%% file: rep.tex
\documentclass{scrartcl}
\makeatletter
\DeclareOldFontCommand{\rm}{\normalfont\rmfamily}{\mathrm}
\DeclareOldFontCommand{\sf}{\normalfont\sffamily}{\mathsf}
\DeclareOldFontCommand{\tt}{\normalfont\ttfamily}{\mathtt}
\DeclareOldFontCommand{\bf}{\normalfont\bfseries}{\mathbf}
\DeclareOldFontCommand{\it}{\normalfont\itshape}{\mathit}
\DeclareOldFontCommand{\sl}{\normalfont\slshape}{\@nomath\sl}
\DeclareOldFontCommand{\sc}{\normalfont\scshape}{\@nomath\sc}
\makeatother
\usepackage[T1]{fontenc}
\usepackage[utf8]{inputenc}
\usepackage[english]{babel} 

\begin{document}

\input{makrot}

\input{kansi}
\pagebreak
\input{2level}

\pagebreak
\bibliography{fysiikka}

\end{document}

%% file: makrot.tex
\newcommand{\be}{\begin{equation}}
\newcommand{\ee}{\end{equation}}
\newcommand{\fr}{\frac}
\newcommand{\pr}{\partial}
\newcommand{\ul}{\underline}
\newcommand{\bea}{\begin{eqnarray}}
\newcommand{\eea}{\end{eqnarray}}

%% file: kansi.tex
\begin{titlepage}
\large

\LARGE
\vspace*{2cm}
\centerline{Exact solutions and magnetic monopoles}
\centerline{in adiabatic three level systems}
\large
\vspace{2cm}
\centerline{Timo Aukusti Laine
\footnote{\large{timo.laine@mlconvex.ai}}}
\vspace{1cm}
\vspace*{2cm}
We investigate the geometrical phase associated to the
Schrödinger equation in a three level
system
in Stimulated Raman Adiabatic Passage (STIRAP). We solve explicitly a dual model,
in which the pulses are
applied in the counter\-in\-tui\-tive and intuitive order.
We show that when the pulse areas are finite, a pair of magnetic
monopoles with opposite charges are created resulting the oscillations of the po\-pulations
on the final states.
The applications of the phase shift include, for example,
phase gates in quantum computing, phase manipulation in quantum cryptography and
phase interactions in quantum interference.
 \\ \\ \\ \\ \\ \\ \\
\end{titlepage}

%% file: 2level.tex
\section{Introduction}

Time-dependent two-state models are widely used in quantum
mechanics. A lot of effort has been made to investigate simple
models describing population transfer at level crossings. The
first prototype model was presented by Laudau \cite{landau}  and
Zener
\cite{zener} where the population was transferred adiabatically.
A review of nonadiabatic corrections and
solvable two-level systems shows that only in a few cases an
exact solution can be found \cite{ka}.

Another class of time-dependent problems arises when we consider
three-level systems. As was pointed out in \cite{ch} and
\cite{tlaine} three-level systems are related to two-level
problems by the SU(2) representation of the rotation
group. Three-level systems are used, for example, to describe
the Stimulated Raman Adiabatic Passage (STIRAP) in quantum optics
as well as neutrino
propagation in the medium in high energy physics.

In this article we study an adiabatic three-level system (STIRAP).
We calculate explicitly a dual model in which the pulses are
applied in the counterintuitive and intuitive order.
We show with the examples how the
 adiabatic phase gets a nontrivial contribution.
The phase manipulation has many applications, for example, in quantum computing and
quantum cryptography. We also show that when the pulse areas are finite, this
corresponds to a case where
magnetic monopoles locate at the origin of the parametric space.

\section{Adiabatic system}

We consider a general time-dependent two-level Schrödinger system
written with the generators of SU(2)

\begin{equation} \label{yhtryhma36}
H_2(t) = \frac{1}{2}
\bar{\sigma}\cdot\bar{R}_2(t).
\end{equation}

\noindent The vector $\bar{R}_2(t)$ is defined as
$\bar{R}_2(t) = \Omega_1(t)
\hat{i}+\Omega_2(t)\hat{k}$
and $\bar{\sigma}$ contains
the Pauli sigma matrices

\begin{equation}
\sigma_1 =
\left[
       \matrix{
          0 & 1
\hspace{-1.5em} \phantom{\frac{\dot{1}}{\sqrt{1}}} \cr
          1 & 0
\hspace{-1.5em} \phantom{\frac{\dot{1}}{\sqrt{1}}} \cr
       }
   \right],
\hspace*{2cm}
\sigma_3 =
\left[
       \matrix{
          1 & 0
\hspace{-1.5em} \phantom{\frac{\dot{1}}{\sqrt{1}}} \cr
          0 & -1
\hspace{-1.5em} \phantom{\frac{\dot{1}}{\sqrt{1}}} \cr
       }
   \right],
\end{equation}

\noindent according to the rule $\sigma_i\sigma_j =
i\epsilon_{ijk}\sigma_k+\delta_{ij}$ $(i,j,k=1,2,3)$.
The state vector of the system Eq.(\ref{yhtryhma36}) is
$|\Psi\rangle$.
By making an unitary transformation $U$
to the state
 $|\Psi \rangle$, $|\tilde{\Psi} \rangle = U|\Psi\rangle =
U(|a_1\rangle+|a_2\rangle)$,
we diagonalize
the Hamiltonian
$U^{\dagger} H_2U = D$
with the eigenvalues
$\pm R_2/2$.
In the adiabatic space Eq.(\ref{yhtryhma36}) transforms as

\begin{equation} \label{yhtryhma37}
H_2^{ad}(t) = (D-iU^{\dagger}\partial_t U)(t) =
\frac{1}{2}
\bar{\sigma}\cdot\bar{R}_2^{ad}(t),
\end{equation}

\noindent where
$\bar{R}_2^{ad}(t) = -
\dot{\phi}(t)\hat{j}+R_2(t)\hat{k}$.
Here we have used the notations
$R_2^2(t) = \Omega_1^2(t)+\Omega_2^2(t)$
and
$\tan \phi(t) = \Omega_1(t)/\Omega_2(t)$.

We apply the result to the STIRAP problem. It consists of
three-levels which are coupled in a sequence
$1\rightarrow 2\rightarrow 3$, and the states 1 and 3 are assumed to
be in resonance \cite{hi}. The Hamiltonian which describes the system is

\begin{equation} \label{hamm3}
 H_3(t) = \bar{J}\cdot \bar{R}_3(t).
\end{equation}

\noindent The vector $\bar{R}_3(t)$ is defined as $\bar{R}_3(t) =
\lambda_1(t)
\hat{i}+\lambda_2(t)\hat{k}$ and $\bar{J}$
contains the modified generators of the rotation
group SO(3)

\begin{equation}
J_1 =
     \left[
       \matrix{
          0 & 1 & 0 \cr
          1 & 0 & 0 \cr
          0 & 0 & 0 \cr
       } \right],
\hspace*{2cm}
J_3 =
     \left[
       \matrix{
          0 & 0 & 0 \cr
          0 & 0 & 1 \cr
          0 & 1 & 0 \cr
       } \right],
\end{equation}

\noindent according the rule $[J_i,J_j] = i\epsilon_{ijk} J_k$
$(i,j,k=1,2,3;i\not= j)$.
The state vector of Eq.(\ref{hamm3}) is $|\psi\rangle =
b_1|1\rangle +
b_2|2\rangle +
b_3|3\rangle$. The question is to find $|b_3(\infty)|^2$ with the
initial condition $|b_1(-\infty)|^2 = 1$. The couplings have to
satisfy $\lim_{t\rightarrow -\infty}\limits
\frac{\lambda_1(t)}{\lambda_2(t)} = 0$ and
$\lim_{t\rightarrow \infty}\limits
\frac{\lambda_2(t)}{\lambda_1(t)} = 0$
when the pulses are applied in the counterintuitive
case.
The SU(2)
representation of the rotation group SO(3)
allows us to present
the current three-level problem as a two-level system with the Hamiltonian
Eq.(\ref{yhtryhma37}) and the relations $\bar{R}_2(t) =
\bar{R}_3(t)$. The result is

\begin{eqnarray}
  b_1 &=& -\sin\phi(a_1a_2^*+a_1^*a_2)+\cos\phi(|a_1|^2-
|a_2|^2),\\
  b_2 &=& -(a_1a_2^*-a_1^*a_2),\\
  b_3 &=& -\cos\phi(a_1a_2^*+a_1^*a_2)-\sin\phi(|a_1|^2-
|a_2|^2).
\end{eqnarray}

\noindent In the counterintuitive case
the corresponding initial condition in the two-level system is
$|a_1(-\infty)|^2 = 1$
which also determines the final
population  $|b_3(\infty)|^2 = |2|a_1(\infty)|^2 -1|^2$.

\section{Exact adiabatic three-level solution}

The couplings $\lambda_1$ and $\lambda_2$ are defined so that

\begin{eqnarray}
  \lambda_1(t) &=& f_1(t)g(t),\\
  \lambda_2(t) &=& f_2(t)g(t),
\end{eqnarray}

\noindent and the corresponding values in the eigenspace become

\begin{eqnarray}
  R_2^2 &=& (f_1^2+f_2^2)g^2,\\
  \dot{\phi}  &=& \frac{\dot{f}_1f_2-f_1\dot{f}_2}
   {f_1^2+f_2^2}.
\end{eqnarray}

\noindent Especially when choosing the functions $f_1$ and $f_2$ in such
a way that

\begin{equation}
  f_1^2+f_2^2 = 1,
\end{equation}

\noindent the parameters $R_2$ and $\dot{\phi}$
separate from each other in the adiabatic space, i.e. \hspace*{-0.2cm}$R_2$ is
a function of $g$ only and $\dot{\phi}$ depends on $f_1$ and
$f_2$ only.
By choosing suitable functions for $g$ we can tune the
behavior of the system
in the adiabatic space.

\subsection{Counterintuitive solution}

As an example of a separated counterintuitive three-level solution we
consider a system whose couplings are

\begin{eqnarray}
  &&\lambda_1(t) = \Omega_1(t) = A\sqrt{\frac{1+\tanh(t/T)}{2}}
\hspace*{1mm}{\rm sech }(t/T) ,
\phantom{\frac{A}{A}} \\
  &&\lambda_2(t) = \Omega_2(t) = A\sqrt{\frac{1-\tanh(t/T)}{2}}
\hspace*{1mm}{\rm sech }(t/T).
\phantom{\frac{A}{A}}
\end{eqnarray}

\noindent $A$ and $T$ are scaling parameters.
In the adiabatic space these couplings transform to the parameters

\begin{eqnarray}
    R_2(t) &=& A\hspace*{1mm}{\rm sech }(t/T),\\
  \dot{\phi}(t) &=&
\frac{1}{2T}\hspace*{1mm}{\rm sech }(t/T).
\end{eqnarray}

\noindent Now the Hamiltonian system of Eq.(\ref{hamm3}) is trivially solved.
A full exact solution is

\begin{eqnarray} \label{yhtryhma33}
b_1(t) &=&
              \tanh\eta\sin\phi\sin(2I)
         +\cos\phi[\hspace*{1mm}{\rm sech }^2\eta+
 \tanh^2\eta\cos(2I)],\\
b_2(t) &=&
          -i2\tanh\eta\hspace*{1mm}{\rm sech }\eta
\sin^2(I),\\
b_3(t) &=& \label{yhtryhma333}
\tanh\eta\cos\phi\sin(2I)
         -\sin\phi[\hspace*{1mm}{\rm sech }^2\eta+
 \tanh^2\eta\cos(2I)],
\end{eqnarray}

\noindent where $I(t) = AT\cosh\eta \hspace*{1mm} {\rm artan}
[\exp(t/T)]$ and $\sinh\eta = 1/(2AT)$.
The final populations become

\begin{eqnarray}
 b_1(\infty) &=&
          \tanh\eta\sin(\pi AT\cosh\eta)
\hspace{-1.5em} \phantom{\frac{\dot{1}}{\sqrt{1}}},\\
b_2(\infty) &=&
          -i {\displaystyle \frac{2\tanh\eta}{\cosh\eta}}
\sin^2\Bigl [ \Bigl (
{\displaystyle \frac{\pi AT}{2}} \Bigr ) \cosh\eta \Bigr ]
\hspace{-1.5em} \phantom{\frac{\dot{1}}{\sqrt{1}}},\\
b_3(\infty) &=&
         -1+{\displaystyle
\frac{\sinh^2\eta[1-\cos(\pi AT\cosh\eta)]}
{\cosh^2\eta}}
\hspace{-1.5em} \phantom{\frac{\dot{1}}{\sqrt{1}}}.
\end{eqnarray}

\noindent We notice that
 $|b_3(\infty)|^2$ depends on $AT$ in a
polynomial way. In the corresponding 2-level system, these oscillations
relate to the Rabi cycle. When

\begin{equation}
  (4n)^2 - (2AT)^2 = 1,
\end{equation}

\noindent and $n > 0$ is an integer, then $|b_3(\infty)|^2 = -1$. Oscillations continue
to the infinity.
Similar numerical results were provided in \cite{tlaine}.

\hspace{1cm}

\noindent We make two remarks here:

\begin{itemize}
  \item If the $b_1(-\infty)$ is a real positive number, then the intermediate state $b_2(t)$
  is populated by the phase only, and the final state, $b_3(t)$
  will be a negative number with no phase.
  During the transition the state $b_3(t)$ obtains an additional
  geometrical phase, $\exp(i\pi)$.
  \item Generally, the state is a complex number with some phase. However, in
  the counterintuitive process with the
  initial condition
  of no phase, the Im$(b_1(t))$, Re$(b_2(t))$, and Im$(b_3(t))$ are never populated.
  From this it follows that
  if we have an initial condition where both real and imaginary parts are populated,
  the solution can be broken into two separate equations
  and the solution can be calculated separately.
\end{itemize}

\subsection{Exponential pulses}

As a comparison we state the result of the exponential pulses
in which the pulse areas not finite \cite{tlaine}. The couplings
are

\begin{eqnarray}
  &&\lambda_1(t) = A( 1 + e^{-t/T})^{-1/2},
\phantom{\frac{A}{A}} \\
  &&\lambda_2(t) = A( 1 + e^{t/T})^{-1/2}.
\phantom{\frac{A}{A}}
\end{eqnarray}

\noindent With the initial condition $b_1(-\infty) = 1$, the
population of the final state is

\begin{equation} \label{lopputulos}
 b_3(\infty) = -1 + {\rm sech }^2(\pi AT).
\end{equation}

\noindent There are two notes:

\begin{itemize}
  \item The population on the level 3 does not oscillate. The Hamitonian degenerates
  and the zero eigenvalue dominates.
  \item The population on the level 3 has the same phase shift $\exp(i\pi)$
  in both cases, i.e. when the pulse areas are finite and infinite.
\end{itemize}

\subsection{Intuitive order solution}

Using the same technique presented before,
it is also possible to calculate the
case when the pulses are applied in the
intuitive order, i.e.

\begin{eqnarray}
  &&\lambda_1(t) = A\sqrt{\frac{1-\tanh(t/T)}{2}}
\hspace*{1mm}{\rm sech }(t/T) ,
\phantom{\frac{A}{A}} \\
  &&\lambda_2(t) = A\sqrt{\frac{1+\tanh(t/T)}{2}}
\hspace*{1mm}{\rm sech }(t/T).
\phantom{\frac{A}{A}}
\end{eqnarray}

\noindent The calculation is similar as earlier and we just state the
result which is

\begin{eqnarray} \label{yhtryhma42}
b_1(t) &=& \sin\phi\cos(2I)
         +\cos\phi\tanh\eta\sin(2I),\\
b_2(t) &=& -i\hspace*{1mm}{\rm sech }\eta
\sin(2I),\\
b_3(t) &=&
\cos\phi\cos(2I)
        -\sin\phi\tanh\eta\sin(2I).
\end{eqnarray}

\noindent When the time is infinite this becomes

\begin{eqnarray} \label{yhtryhma43}
b_1(\infty) &=&
\tanh\eta\sin(\pi AT\cosh\eta),\\
b_2(\infty) &=&
          -i\hspace*{1mm} {\rm sech}\eta\sin(\pi AT\cosh\eta),\\
b_3(\infty) &=&
\cos(\pi AT\cosh\eta).
\end{eqnarray}

\noindent We note that $|b_3(\infty)|^2$ is now pure oscillations. There are
two special cases

\begin{eqnarray}
  &&|b_3(\infty)|^2 = 1, \hspace*{2cm} (2n)^2 - (2AT)^2 = 1, \\
  &&|b_3(\infty)|^2 = 0, \hspace*{2cm} \Bigl[2(n-\frac{1}{2})\Bigr]^2 - (2AT)^2 = 1,
\end{eqnarray}

\noindent and $n > 0$ is an integer.

\hspace{1cm}

We conclude that we have been able to solve
both of the cases,
counterintuitive and intuitive order,
by using the same structure
of the pulses. We call this as a dual model.

\section{Three dimensional system}

We consider a time- and space-dependent Schrödinger equation

\begin{equation} \label{schro3d}
  i\hbar\frac{\partial}{\partial t} \Psi(R,t) =
  (T+V)\Psi(R,t),
\end{equation}

\noindent where $T$ is a kinetic energy operator and $V$ is a potential.
A well-known numerical
solution approach
is a Split operator -method, in which the solution is written in an exponential form.
The operators generally do not commute, but using the Baker-Hausdorff formula, one can show
that within a small time interval, $\Delta T$, the error will be of order $dt^2$.
When the time interval is small enough, the numerical result will be accurate.
The Split operator -method
solution of Eq.(\ref{schro3d}) can be written as

\begin{eqnarray} \label{splitratk1}
  \Psi(R,t+\Delta t) &=&
  e^{-iV\Delta t/(2\hbar)}
    e^{-iT\Delta t/\hbar}e^{-iV\Delta t/(2\hbar)}
  \Psi(R,t) \\
  &=& \hat{O}(T,V)\Psi(R,t).
\end{eqnarray}

\noindent The operator $T$ is solved in the Fourier space.
Let's assume we have a normalized minimum wave function

\begin{equation}
   \Psi(x) = [2\pi(\Delta x)^2]^{-1/4}
        \exp\Bigl[-\frac{(x-\langle x\rangle)^2}{4(\Delta x)^2}
                  +\frac{i\langle p\rangle
x}{\hbar}\Bigr],
\end{equation}

\noindent which propagates in an harmonic oscillator potential

\begin{equation}
 V \sim  \frac{1}{2}m\omega^2(x-x_0)^2.
\end{equation}

\noindent According to the Eq.(\ref{splitratk1})
the population on the level is interchanged between the real and imaginary parts of the state
and the wave probability remains the same.

Now we consider the Hamiltonian which
includes the couplings $\lambda_1$ and $\lambda_2$

\begin{equation} \label{schrode2}
 H_{3D}
 =
     \left[
       \matrix{
          T_1(p)+V_1(x) & \lambda_1(t) & 0 \cr
          \lambda_1(t) & T_2(p)+V_2(x) & \lambda_2(t)
\cr
          0 & \lambda_2(t) & T_3(p)+V_3(x) \cr
       } \right],
\end{equation}

\noindent and the Schrödinger equation is

\begin{equation}
  i\hbar\frac{\partial}{\partial t} \Psi(R,t) =
  (T+V+\lambda)\Psi(R,t).
\end{equation}

\noindent The Split operator -solution is then

\begin{eqnarray} \label{splitratk2}
  \Psi(R,t+\Delta t) &=& e^{-i(V+\lambda)\Delta t/(2\hbar)}
    e^{-iT\Delta t/\hbar}
    e^{-i(V+\lambda)\Delta t/(2\hbar)}
  \Psi(R,t) \\
  &=& \hat{F}(\lambda)\hat{O}(T,V)\hat{F}(\lambda)\Psi(R,t),
\end{eqnarray}

\noindent where $\hat{F}(\lambda) = \exp(-i\lambda\Delta t/(2\hbar))$.
 What this equation states, is that in a small time interval, we first make a small
STIRAP process step, $\hat{F}(\lambda)$,
from level 1 to 3, then interchange the population between the real and
imaginary parts on all 3 states, $\hat{O}(T,V)$, and then make
another small STIRAP process step, $\hat{F}(\lambda)$, from level 1 to 3.
In this solution method these two processes are not connected; one transfers
the population between the states and another moves the population within the state.
Also, as it was
stated in the previous section, in the STIRAP process the full complex wave
equation can be separated into two different equations
which can be solved separately. Altogether this means that under the
STIRAP process, the full wave function which has some phase, is transferred to the final state in a
same manner as in our simple exact solution, as a result that the wave function
has obtained an additional geometric phase $\exp (i\pi)$.
Depending on the potentials of the states,
the wave functions can change their shapes during the transition.

\subsection{Phase manipulation}

We consider two consecutive STIRAP processes. The first
counterintuitive STIRAP process
transfers the population from level 1 to 3, and the level 3 obtains the phase shift.
Then another STIRAP is applied in the reverse order (intuitive order)
and the population is transferred back from state 3 to 1. If the delay between the
pulses is $\Delta T$, the geometric phase on state 1 is shifted by $\Delta T$. By changing
the value of $\Delta T$, one can rigorously control the geometric phase on the state 1.

\hspace{1cm}

The phase manipulation has applications, for example, in the following areas:

\begin{itemize}
  \item One-qubit phase gate in quantum computing.
  \item Phase manipulation in quantum cryptography.
  \item Phase Conjugated Mirror (PCM) type of applications.
  \item Magnetic charge ($\pm$) annihilation (analogy to electron-positron annihilation).
  \item In the study of quantum interference.
\end{itemize}

\noindent All these phenomena make use of the phase shift and a complete two-way population
transfer between the states.

\subsection{Hadamard gate}

Another popular phase gate in quantum computing is the Hadamard gate. It has
the Hamiltonian

\begin{equation} \label{hadamard}
H(t) =
     \left[
       \matrix{
          0 & 0 & \lambda_{10}(t) & 0 \cr
          0 & 0 & \lambda_{11}(t) & 0 \cr
          \lambda_{10}(t) & \lambda_{11}(t) & 0 & \lambda_2(t) \cr
          0 & 0 & \lambda_2(t) & 0 \cr
       } \right],
\end{equation}

\noindent and the state vector is $|\psi\rangle =
b_{10}|10\rangle +
b_{11}|11\rangle +
b_2|2\rangle +
b_3|3\rangle$. We use the notations

\begin{eqnarray}
  &&\lambda_{10}(t)  = c_{10}\lambda_1(t), \\
  &&\lambda_{11}(t)  = c_{11}\lambda_1(t), \\
  && |1\rangle = c_{10}|10\rangle + c_{11}|11\rangle,
\end{eqnarray}

\noindent where $c_{10}$ and $c_{11}$ are constants. When defining the constants
to have a relation

\begin{equation}
   c_{10}^2 +c_{11}^2 = 1,
\end{equation}

\noindent the Hamiltonian Eq.(\ref{hadamard}) reduces to the 3-dimensional
Hamiltonian Eq(\ref{hamm3}). Now
the same exact solution of Eqs.(\ref{yhtryhma33})-(\ref{yhtryhma333})
can be applied to find exact solutions to this system also.

One finding is that when the pulse areas are finite,
the Rabi cycle is also at present in the Hadamard gate
and the final populations on the levels oscillate. The Rabi cycle
has crucial importance in quantum computing.

\section{Monopoles in adiabatic three-level system}

We consider two approaches to
show the existence of the monopoles in the STIRAP system
when the pulse areas are finite. The first formalism is commonly
used in quantum optics while the second approach uses the notation of quantum field theory.

\subsection{Approach A}

When the pulse areas of the counterintuitive STIRAP system are
finite,
a soliton with a constant flux can be found.
We show this by using the Berry's
adiabatic phase and the group SU(2). The definition of the
Berry's
phase is \cite{dr}

\begin{equation}
 \gamma_m(C) = \oint_C d\bar{R}\cdot\bar{A}(\bar{R}) =
\int_S d\bar{S}\cdot\bar{V},
\end{equation}

\noindent where $\bar{R}$ describes the slowly varying parameters in time
and $\bar{A}(\bar{R})$ is a vector potential of the magnetic
field. In the second
equality we have used Stoke's Law and defined $\bar{V} =
\nabla\times\bar{A}$ which is the flux associated to the
magnetic field. Writing
the Berry's
phase in the form which is manifestly independent of the phase
of state $|m,\bar{R}\rangle$, we get

\begin{equation}
  \gamma_m(C) = -\int_S d\bar{S}\cdot\bar{V}_m,
\end{equation}

\noindent where

\begin{equation} \label{vm}
 \bar{V}_m(\bar{R}) =
{\rm Im} \sum_{n\not= m}\limits
\frac{\langle m,\bar{R}|\nabla_R H|n,\bar{R}\rangle \times
\langle n,\bar{R}|\nabla_R H|m,\bar{R}\rangle}
{(E_m(\bar{R})-E_n(\bar{R}))^2}.
\end{equation}

\noindent When the eigenvalues cross, a field source is at present.
Evaluating Eq.(\ref{vm}) for a Hamiltonian Eq.(\ref{yhtryhma36}) we
get

\begin{equation} \label{poten}
  \bar{V}_{+}(\bar{R}_2) = +\frac{1}{2}\frac{\hat{R}_2}{R_2^2},
\hspace*{2cm}
  \bar{V}_{-}(\bar{R}_2) = -\frac{1}{2}\frac{\hat{R}_2}{R_2^2},
\end{equation}

\noindent and the degeneracy exists when $R_2(-\infty) = R_2(\infty) = 0$.
The Berry's phase becomes

\begin{equation}
  \gamma_{\pm} =
\mp \frac{1}{2} \Delta\Omega,
\end{equation}

\noindent where $\Delta\Omega$ is the solid angle subtended by the closed
path as seen from the place of degeneracy, $R_2 = 0$.
The $\pm$ refers to the direction in which the line integration
is traversed. Equations
(\ref{poten})
equal with the Wu-Yang magnetic monopole of
strength
$\pm 1/2$ located at the origin in the parameter space \cite{wy}. The
total flux of the monopole is $\Phi = \pm 2\pi$.

\subsection{Approach B}

The transformation from the two-level system Eq.(\ref{yhtryhma36})
to the adiabatic Hamiltonian Eq.(\ref{yhtryhma37}) can be seen as a
special case of a local gauge transformation

\be \label{gfield}
 D_t\psi^a \equiv \partial_t\psi^a+
\epsilon^a_{\phantom{a}bc}A_t^b\psi^c = 0.
\ee

\noindent $D_t$ is a covariant derivative, and $A_t^i$ is
a gauge field with the non-Abelian group
SU(2),  $\epsilon^a_{\phantom{a}bc}A_t^b\psi^c =
i(H_2(t)\Psi)^a$.
One special configuration of $\Psi$ defines a supervacuum and by
using the SU(2) rotations we can go to another vacuum, which we
call a normal vacuum. We look for soliton solutions
which are topologically stable. The energy must be finite and
this is achieved by choosing the proper boundary conditions for
$A^a_t$.

We show the existence
of a string like soliton in the Schrödinger equation,
Eq.(\ref{hamm3})

\be \label{schrode100}
  D_t\Psi = (\partial_t + iA_t)\Psi = 0,
\ee

\noindent where $A_t=H_3(t)$.
Here $t$ is some arbitary parameter which parametrizes the space.
First we look the symmetries of Eq.(\ref{schrode100}). Clearly it is
not any more locally gauge invariant. After an infinitesimal
rotation we get diagonal matrix elements, which do not belong to
the group SO(3).
The vacuum consists of three scalar fields, and
it has the SO(3) symmetry because the norm is conserved, $
(\psi^1)^2+(\psi^2)^2+(\psi^3)^2 = 1$.
The supervacuum is determined by fixing the
initial condition, $\psi^1(-\infty)^2 = 1$.
The analogy in the field theory is the Higgs field, which
specifies the vacuum state.
The
symmetry of the supervacuum is thus U(1), ${\cal M}_0 = S^1$,
which is needed for a soliton carrying a magnetic charge.
The unbroken vacuum has the symmetry SO(3)/U(1).

Again we are
looking for a finite energy solution. The eigenvalues of $H_3$
define the energy of the system
 $\omega_0(t) = 0,$
$\omega_{\pm}(t) = (\lambda_1^2(t)+\lambda_2^2(t))^{1/2}$.
For a finite energy solution the eigenvalues must vanish
$\omega_{\pm} \rightarrow 0$, when $t \rightarrow \pm \infty$.
It follows that the gauge
field also vanishes, $A_t \rightarrow 0$ when $t \rightarrow \pm
\infty$. The manifold of the points at infinity is thus ${\cal
M}_{\infty}$ = SO(3)/U(1). In order to have a nontrivial
solution we need a well defined map from ${\cal M}_{\infty}$
into ${\cal M}_0$. The first homotopy class of the group SO(3),
$\pi_1(SO(3)) = Z_2$, assures that a path which forms a circle
belongs to the ${\cal M}_{\infty}$. The map is then $S^1
\rightarrow S^1$.
All such maps are cylidrically symmetric and characterized by
integers, i.e.\hspace*{-0.2cm}
 $ \Psi(r\rightarrow \infty) \rightarrow \exp^{-in\theta}.$
We look for a solution in the form
 $ \Psi = f(\rho)\exp^{-in\theta},$
where $f(\rho) \rightarrow 1$ when $r \rightarrow \infty$ and
$f(\rho)$ vanishes at the origin.
From Eq.(\ref{schrode100}) we get
 $ A_t \rightarrow n\partial_t\theta,$
when
$r\rightarrow\infty.$
We have a string like soliton, whose
 magnetic field is

\be
  \int \bar{B}\cdot d\bar{S} =
\oint_{r\rightarrow \infty} A_t dt = n\Delta\theta,
\ee

\noindent with the flux $\Psi = 2\pi n$.

\subsection{Monopole confinement}

In the case of intuitive order pulse system, the supervacuum does not respect any symmetries
and a constant flux tube does not exist. For a counterintuitive
pulse system, the supervacuum is $S^1$ invariant under the
coupling $\lambda_2(t)$. The system admits then a constant
magnetic flux tube. When the coupling $\lambda_1(t)$ is switched
on, the symmetry of the supervacuum is destroyed. Physically
this looks very odd, since one would expect magnetic flux to be
conserved. One explanation of the breaking up of
double tubes is that a pair of magnetic monopoles with
opposite magnetic charges are created. The supervacuum does not exist
when the pulse areas are infinite.

\section{Conclusions}

In conclusion, we have shown an approach to obtain exact
solutions to adiabatic three-level systems. In the adiabatic space the Hamiltonian becomes solvable.
In particular, we calculated explicitly a dual model in a three-level system
(STIRAP)
where the pulses were
applied in the counterintuitive and intuitive order.
We
showed that when the eigenvalues crosses at the infinity,
 a pair of magnetic monopoles are created.
Interesting is that the monopoles and Rabi cycle are not at present when the pulse
areas not finite.
Additional phase is added in both cases, but the population on level 3 gets
oscillations when the pulse areas are finite.

When two STIRAP processes are applied in a row; first the counterintuitive process and
then the intuitive order process, the population is transferred back to the initial state.
The phase of the new state can be well controlled.
There are many application areas where the phase shift can have a significant impact.
Some of these are, for example,
phase gates in quantum computing, phase manipulation in quantum cryptography and
the phase interactions in quantum interference.